\documentclass[onecolumn,preprint,superscriptaddress]{revtex4-2}

\usepackage{mathrsfs}
\usepackage{amsmath}
\usepackage{amssymb}
\usepackage{graphicx}
\usepackage{float}
\restylefloat{table}
\usepackage[pdftex]{hyperref}
\usepackage{float}
\usepackage{subfigure}
\usepackage[pdftex]{color}
\usepackage{ulem} %pra riscar correções no texto%
\begin{document}

\title{Electron Confinement study in a double quantum dot by means of Shannon Entropy Information\footnote{Version accepted for publication in Phys. B (Amsterdam, Neth.) (2024). \href{https://doi.org/10.1016/j.physb.2024.415692}{{\color{blue}Click here to access.}}}.}

\author{W. S. Nascimento}
\email{wallassantos@gmail.com}
\affiliation{Instituto de F\'{\i}sica, Universidade Federal da Bahia, Campus Universit\'ario de Ondina, 40170-115, Salvador, BA, Brazil}

\author{A. M. Maniero}
\email{angelo.maniero@ufob.edu.br}
\affiliation{Centro das Ci\^encias Exatas e das Tecnologias, Universidade Federal do Oeste da Bahia, 47808-021, Barreiras, BA, Brazil}

\author{F. V. Prudente}
\email{prudente@ufba.br}
\affiliation{Instituto de F\'{\i}sica, Universidade Federal da Bahia, Campus Universit\'ario de Ondina, 40170-115, Salvador, BA, Brazil}

\author{C. R. de Carvalho}
\email{crenato@if.ufrj.br}
\affiliation{Instituto de F\'{\i}sica, Universidade Federal do Rio de Janeiro, Rio de Janeiro, 21941-972, RJ, Brazil}

\author{Ginette Jalbert}
\email{ginette@if.ufrj.br}
\affiliation{Instituto de F\'{\i}sica, Universidade Federal do Rio de Janeiro, Rio de Janeiro, 21941-972, RJ, Brazil}

%\date{\today }

\begin{abstract}

In this work, we use the Shannon informational entropies to study an electron confined in a double quantum dot; we mean the entropy in the space of positions, $S_r$, in the space of momentum, $S_p$, and the total entropy, $S_t = S_r+S_p$. We obtain  $S_r$, $S_p$ and $S_t$ as a function of the parameters  $A_2$ and $k$ which rules the height and the width, respectively, of the internal barrier of the confinement potential. We conjecture that the entropy $S_r$ maps the degeneracy of states when we vary $A_2$ and also is an indicator of the level of decoupling/coupling of the double quantum dot. We study the quantities $S_r$ and $S_p$ as measures of delocalization/localization of the probability distribution. Furthermore, we analyze the behaviors of the quantities $S_p$ and $S_t$ as a function of $A_2$ and $k$. Finally, we carried out an energy analysis and, when possible, compared our results with work published in the literature. \\  


\end{abstract}

\maketitle

\pagebreak

\section{Introduction}

Quantum systems under confinement conditions present notable properties and are studied over a wide range of situations~\cite{sabin2009v57-58, sen2014electronic}. From the theoretical point of view one finds different approaches and methods to treat these systems, by computing the electronic structure of atoms, ions and molecules under the confinement of a phenomenological external potential in the presence (or not) of external fields. For instance, analytical approximations for two electrons in the presence of an uniform magnetic field under the influence of a harmonic confinement potential representing a single quantum dot (QD)~\cite{merkt1991, wagner1992, quiroga1993}, or a quartic one corresponding to a double QD~\cite{burkard1999}; in this last case the presence of an additional laser field is done through the electron effective mass~\cite{carvalho2003}; or different numerical methods of calculation related to the concern with the accuracy of  describing the electron-electron interaction in (artificial) atoms, molecules and nanostructures such as the Hartree approximation~\cite{pfannkuche1993, creffield2000},  the Hartree-Fock computation~\cite{szafran2004, thompson2005, olavo2016} or the full conﬁguration interaction method (Full CI)~\cite{jung2003, thompson2005, olavo2016} among others. Besides, initially the study of confined quantum systems involved the study of the electronic wave function in an atom, or ion, inside a box, whose walls could be partially or not penetrable, and whose description led to the use of different phenomenological potentials~\cite{froman1987, connerade2000}. In the case of QD's, the choice of potential profiles has usually involved a harmonic profile~\cite{olavo2016, maniero2020a, maniero2020b, maniero2021}, or an exponential one, to take into account the finite size of the confining potential well~\cite{Adamowski2000, Xie2003, maniero2023}; or a combination of both~\cite{duque-2023-magnetico, duque2023-laser}. Recently, we have analyzed the behavior of two electrons in a double QD with different confinement profiles, and under the influence of an external magnetic field, aiming at interest in fundamental logical operations of quantum gates~\cite{maniero2023}.

On the other hand, the comprehension of the properties of confined quantum systems is related to the choice of what physical quantities are computed and analysed; in the case of QDs  one finds, for instance, the computation of linear and nonlinear absorption coefficients, refractive index, and harmonics generation susceptibilities~\cite{Sargsian2023}, as well as exchange coupling, electron density function and electronic spatial variance~\cite{maniero2020a, maniero2020b}. Although the mathematical basis of information theory was established a long time ago~\cite{nyquist1928, hartley1928, shannon1948a}, only recently informational entropy has been used as an alternative to  the study of the properties of confined quantum systems~\cite{Mukherjee2016, nascimentocoulomb-et_al2021, estanon_etal2020, saha-jose2020, salazar2021, Santos2022} and in particular QDs~\cite{liu2023,PhysRevA.105.032821}.

The present work aims to use Shannon informational entropy as a tool to study an electron confined in a double quantum dot. We use a confinement potential composed of a harmonic-gaussian symmetric double quantum well function and harmonic functions. More precisely, by manipulating parameters of the double quantum well function we analyze, for example, the level of decoupling/coupling between neighboring quantum wells. This treatment allows us to study the formation of degenerate and non-degenerate states, as well as the phenomenon of electron tunneling. This approach has applications, among other topics, in quantum computing, where as observed by Loss~and~DiVincenzo~\cite{Loss1998-uv} the quantum gate operation of two qubits in a double quantum dot is connected to the decoupling/coupling level between the quantum wells.

Throughout this paper we use atomic units and cartesian coordinate axes. The present paper is organized as follows. In Section 2 our theoretical approach is discussed: in Sec.2.1 the concepts and methodology adopted in this work are presented, in particular the phenomenological confinement potential, whose width, height and coupling are adjusted by different parameters; and in Sec.2.2 the entropy quantities are defined for the sake of completeness. The Sec.3 is also divided in Sec.3.1 and Sec.3.2, where energy and entropies, respectively, are studied as functions of the parameters which rule the potential's height and coupling.

\section{Model and Formulation}

This section presents the concepts and methodology of the calculations used in this work.  In particular,   Subsection~2.1 %\ref{systems_of_interest}%
is dedicated to the presentation of the system formed by an electron confined in
a double quantum dot as the physical problem of interest and in Subsection~2.2 %\ref{information_entropies}%
the informational quantities $S_r$, $S_p$ and $S_t$ are defined.

\subsection{System of interest}\label{systems_of_interest}

\subsubsection{Hamiltonian}

In the present work we study a system formed by an electron confined in a double quantum dot, whose Hamiltonian is
\begin{eqnarray}
\hat H =-\frac{1}{2m_c}\vec\nabla^2+ \hat{V}(x,y,z),
\label{hamiltoniano}
\end{eqnarray}
where $m_c$ is the effective electronic mass and the confinement potential function is given by
\begin{eqnarray}
\hat{V}(x,y,z) = \hat{V}_{DQD}(x) + \hat{V}_{HO}(y) +\hat{V}_{HO}(z) \ .
\label{potencialc}
\end{eqnarray}
The potential function $\hat{V}_{DQD}(x)$ is defined by a harmonic-gaussian symmetric double quantum well function, so that,
\begin{eqnarray}
\hat{V}_{DQD}(x) = V_0 \left[   A_1 \frac{x^2}{k^2} + A_2 e^{-\left(\frac{x}{k}\right)^2}  \right],  
%\ \textrm{with} \ \ A_1>0\ \ \ \textrm{and} \ \ A_2>A_1\ ,
\label{DPQS} 
\end{eqnarray}
with $A_1>0$ and $A_2>A_1$,
where $V_0$ is the depth of the well and $k$ is the parameter that relates the width of the confinement barrier. The parameters $A_1$ adjusts the well to the width of the barrier and $A_2$ the height of the internal barrier adjusting the coupling/decoupling between the wells. The potential functions $\hat{V}_{HO}(y)$ and $\hat{V}_{HO}(z)$ are defined by harmonic functions, that is,
\begin{eqnarray}
\hat{V}_{HO}(y) &=& \frac{1}{2}m_c \omega^2_y y^2
\label{HO(y)}
\end{eqnarray}
and
\begin{eqnarray}
\hat{V}_{HO}(z) &=& \frac{1}{2}m_c \omega^2_z z^2 \ ,
\label{HO(z)}
\end{eqnarray}
where the angular frequencies $\omega_y$ and $\omega_z$ indicate the confinement parameters. 

\begin{figure}[ht]
\centering
%\hspace*{-1.2 cm}
\includegraphics[scale=0.35]{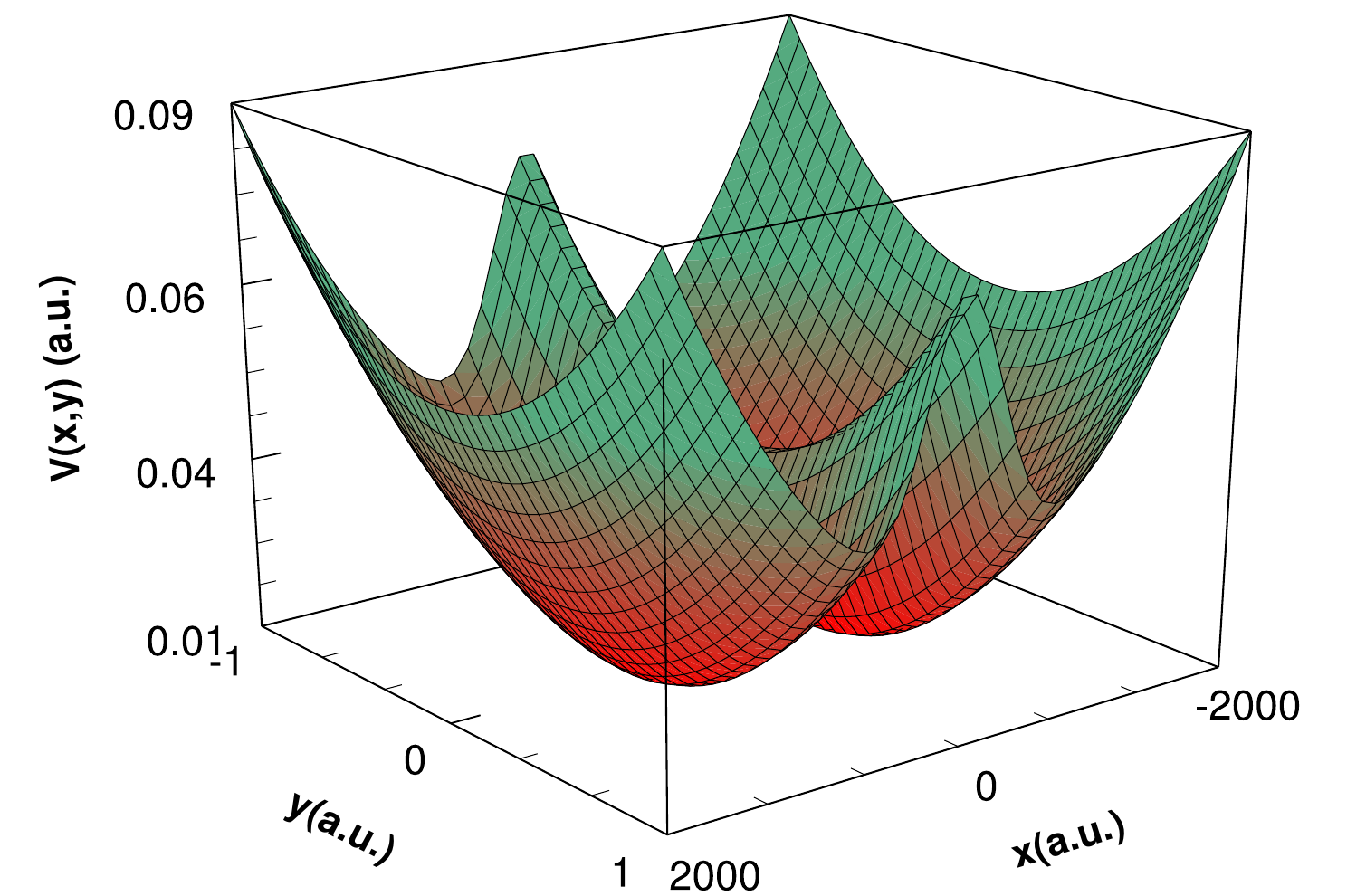}
%\vspace{-1.2 cm}
\caption{Graph of the confinement potential function $\hat{V}(x,y,0)$ for A$_1$~=~0.240, A$_2$~=~5.000, k~=~377.945~a.u., V$_0$~=~0.00839~a.u., $m_c$~=~0.067~a.u. and $\omega_y$~=~1.000~a.u..}               
\label{3Dpotencial}
\end{figure}

In the Fig.~\ref{3Dpotencial} we present the graph of the confinement potential function, $\hat{V}(x,y,0)$, for A$_1$~=~0.240, A$_2$~=~5.000, k~=~377.945~a.u., V$_0$~=~0.00839~a.u., $m_c$~=~0.067~a.u. and $\omega_y$~=~1.000~a.u. In this figure, we observe the form of the potential function that confines the electron in the double quantum dot, including the infinite barriers of confinement and the internal barrier that regulates the decoupling/coupling between the two wells. 

We are interested here in studying the influence of the structure of the double quantum dot on the properties of the system, more precisely, when we vary the parameters $A_2$ and $k$ of the potential function $\hat{V}_{DQD}(x)$. Thus, avoid excitation in the directions $\hat{y}$ and $\hat{z}$ and fixed the situation of spatial confinement in these directions determining the potential functions (\ref{HO(y)}) and (\ref{HO(z)}) with $\omega_y~=1.000$~a.u. and $\omega_z~=1.000$~a.u..
 
\begin{figure}[ht] % Inicia o ambiente de figuras
  \subfigure{ % Começa a incluir a figura fig1.pdf
    \includegraphics[scale=0.36]{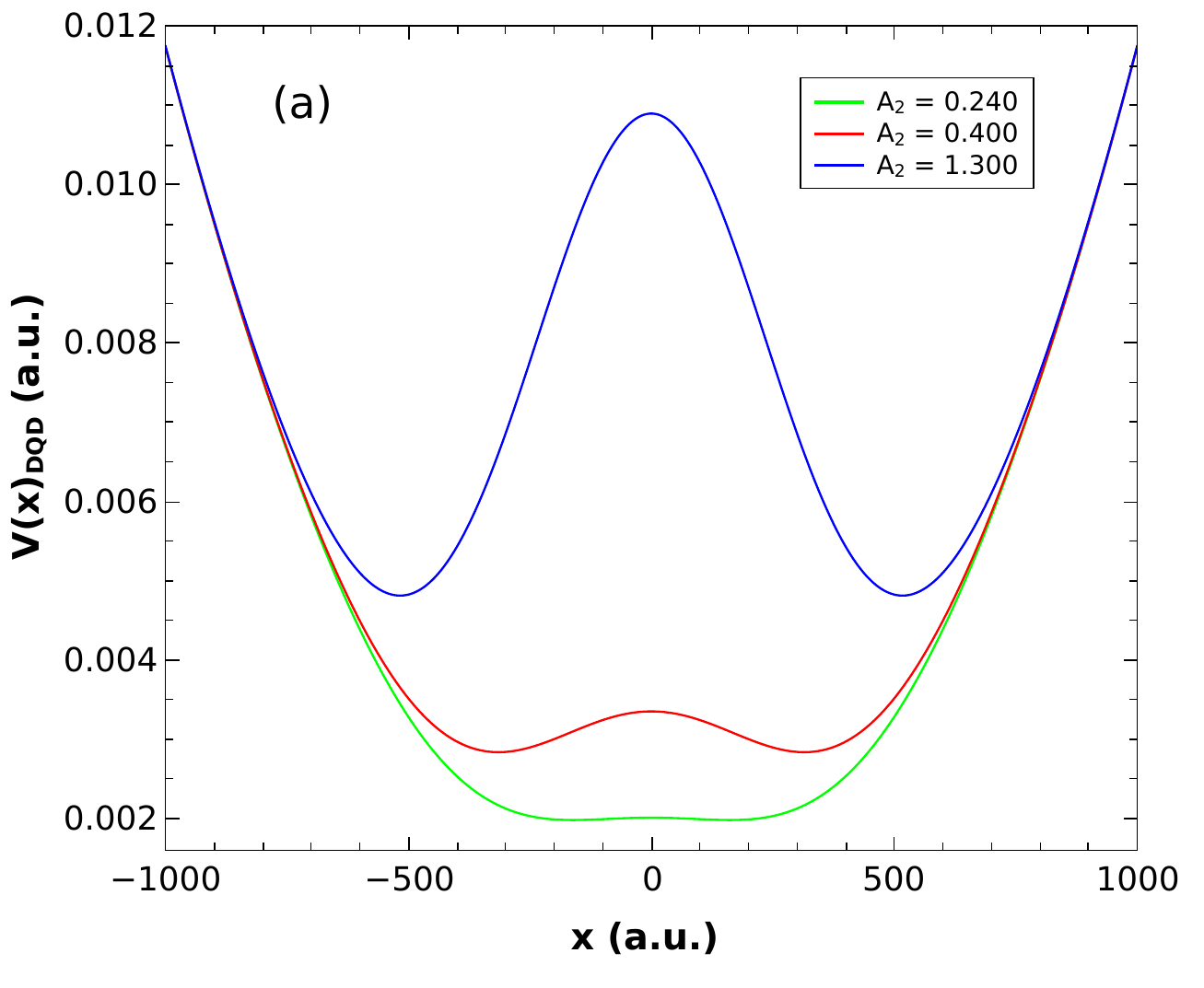}
  } % Termina de incluir a figura fig1.pdf
  \subfigure{ % Começa a incluir a figura fig2.pdf na mesma linha da figura fig1.pdf
    \includegraphics[scale=0.35]{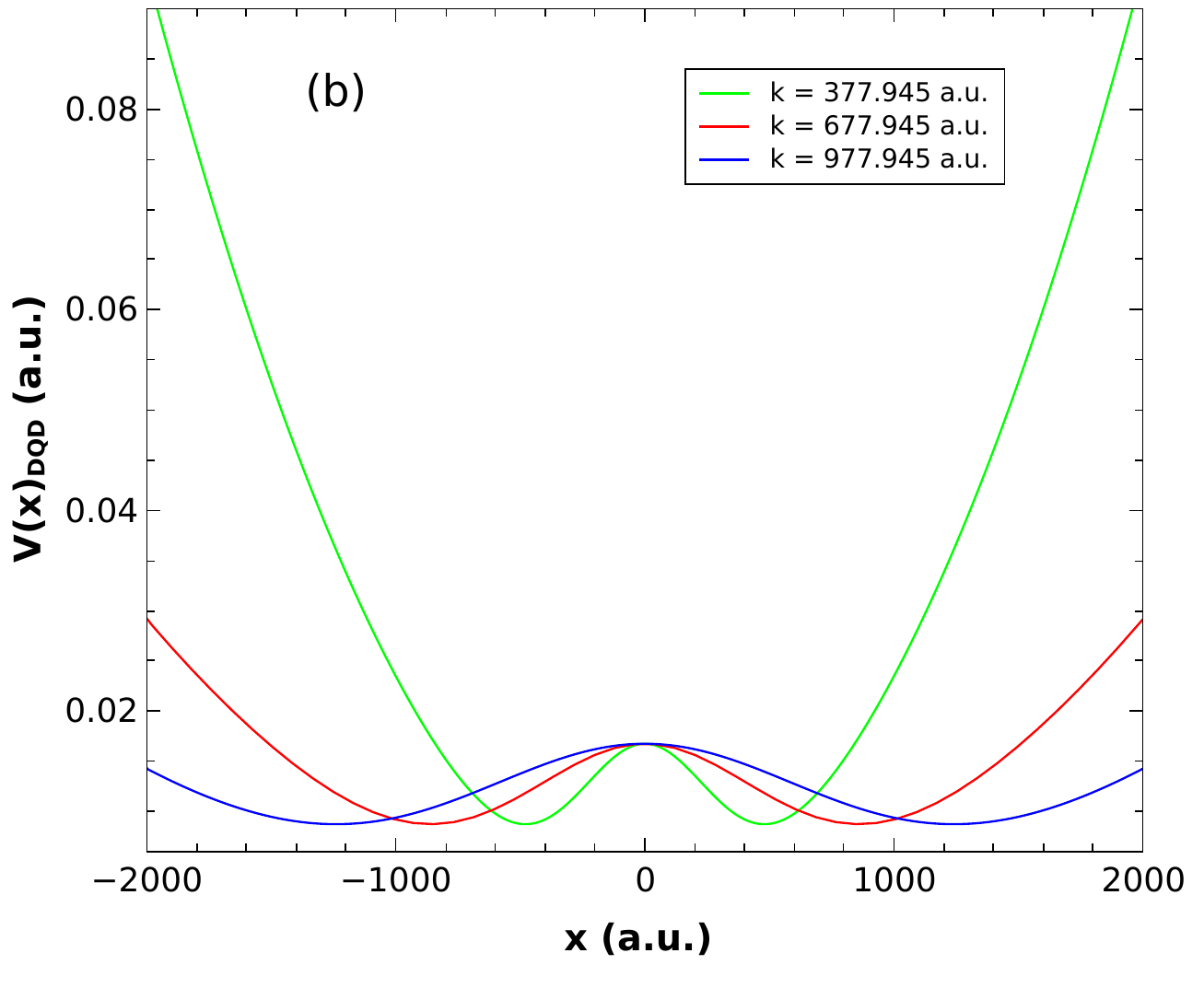}
  } % Termina de incluir a figura fig2.pdf
  \caption{Harmonic-gaussian symmetric double quantum well as a function of $x$ for: (a) different values of $A_2$ with $k=$377.945 a.u., $A_1=$0.200 and $V_0=$0.00838~a.u. and (b) different values of $k$ with $A_2=$0.400, $A_1=$2.000 and $V_0=$0.00838~a.u..}
\label{potencialx}
\end{figure}

In Fig.~\ref{potencialx} we present graphs with the general behavior of the potential function $\hat{V}_{DQD}(x)$ when: (a) we vary the parameter $A_2$ with fixed $k$, $A_1$ and $V_0$ and (b) we change the values of $k$ with fixed $A_2$, $A_1$ and $V_0$. From graph (a) we see that the increase in $A_2$ values increases the level of decoupling between the two wells. Additionally, when the values of $A_1$ and $A_2$ are very close we have approximately one well in $V_{DQD}(x)$. According to graph (b), the increase in $k$ values increases the width of the confinement barrier. The minimum values of $V_{DQD}(x)$ are changed according to variations in $A_2$ or $k$.

 % Fecha o ambiente de figuras

For a deeper understanding of the behavior of the potential function $V_{DQD}(x)$ we present in Fig.~S1 of the supplementary material the situations in which we vary the parameter $A_1$ with $k$, $A_2$ and $V_0$ fixed and where we change $V_0$ with $A_1$, $A_2$ and $k$ fixed.

\subsubsection{Wave functions and probability densities}

Our problem is to solve the Sch\"odinger equation
\begin{eqnarray}
\hat H \psi(x,y,z) = E\psi(x,y,z)
\end{eqnarray}
adopting the Hamiltonian~(\ref{hamiltoniano}). Here, we write the wave function of an electron 
\begin{eqnarray}
\psi(x,y,z)= \psi_x(x)\psi_y(y)\psi_z(z) \
\label{psi}
\end{eqnarray}
in terms of basis functions of the Cartesian anisotropic Gaussian orbitals (c-aniGTO) type centered at $\vec{R}=(X,0, 0)$, that is,
\begin{eqnarray}
\psi_x(x) &=& \sum_\mu N_{\mu} C_{\mu}(x-X_{\mu})^{n_x}\exp[-\alpha_{\mu}(x-X_{\mu})^2] . \label{funcaox} \\
\psi_y(y) &=& N_y \ y^{n_y}\exp[-\alpha_y y^2] \ \label{funcaoy} {\rm and}\\
\psi_z(z) &=& N_z \ z^{n_z}\exp[-\alpha_z z^2]  , \label{funcaoz}
\end{eqnarray}
where $N_{\mu}$, $N_y$ and $N_z$ are the normalization constants, $C_{\mu}$ are the molecular orbital coefficients obtained by diagonalization methods and $X_{\mu}$ is defined in $\pm
k\sqrt{\ln\left( {A_2 / A_1} \right)}$ (minimum values of $\hat{V}_{DQD}(x)$) 
The integers $n_x$, $n_y$ and $n_z$ allow the classification of orbitals, for example, the types $s-$, $p-$, $d-$, ... correspond to $n = n_x + n_y + n_z=$ 0, 1, 2, ...., respectively. 

In Eq.~(\ref{funcaox}), as we did in previous  articles(Refs.~\cite{olavo2016,maniero2019,maniero2020a,maniero2020b,maniero2021,maniero2023}), we have considered two types of exponents in $x$, the first Gaussian exponent $\alpha_x$ has been obtained variationally, that is, minimizing the functional energy in $x$, and the second proportional to the first as being~$\alpha=\frac{3}{2}\alpha_x$. In its turn, in Eqs.~\ref{funcaoy}~and~\ref{funcaoz}, $n_y$ and $n_z$ were taken equal to 0 to avoid excitation in the directions~$\widehat{y}$~e~$\widehat{z}$.  Furthermore, taking $\alpha_y=\alpha_z=\alpha=\frac{1}{2}m_c\omega_y$ and $N_y=N_z=N$, we have  
\begin{eqnarray}
\psi_y(y) &=& N\exp[-\alpha (y)^2] \  \label{funcaoyy} ,\\
\psi_z(z) &=& N\exp[-\alpha (z)^2] \ .\label{funcaozz}
\end{eqnarray}

The wave function $\widetilde{\psi}(p_x, p_y, p_z)$, in momentum space, has been obtained  through a Fourier transform applied to $\psi(x,y,z)$, so that we get
\begin{eqnarray}
\widetilde{\psi}(p_x, p_y, p_z) = \widetilde{\psi_x}(p_x) \widetilde{\psi_y}(p_y) \widetilde{\psi_z}(p_z) ,
\label{ppsi}
\end{eqnarray}
where
\begin{eqnarray}
	\widetilde{\psi_x}(p_x)&=&
	\dfrac{1}{\sqrt{2\pi}}
	\sum_\mu N_\mu C_\mu
		e^{ -\frac{p_x^2}{4\alpha_\mu} -i{p_xX_\mu} }\times\nonumber\\
	&&\sum_{k=0 \atop k\text{ even}}^{n_\mu}
	\binom{n_\mu}{k}
	\Big(\dfrac{p}{2i\alpha_\mu}\Big)^{n_\mu-k}
	\dfrac{\Gamma\big(\dfrac{k+1}{2}\big)}{\alpha_\mu^{(k+1)/2}}  , \label{psipx}\\
	\widetilde{\psi_y}(p_y) &=& \dfrac{N}{\sqrt{2\alpha}}\exp\left(-\dfrac{p_y^2}{4\alpha}\right)  , \label{psipy} {\rm and}\\
	\widetilde{\psi_z}(p_z) &=& \dfrac{N}{\sqrt{2\alpha}}\exp\left(-\dfrac{p_z^2}{4\alpha}\right)  . \label{psipz}
\end{eqnarray}

The probability density in the position space is defined as usual as
\begin{eqnarray}
\rho(x,y,z) &=& \rho_x(x)\rho_y(y)\rho_z(z) \nonumber \\
&=& \vert \psi_x(x) \vert^2 \vert \psi_y(y) \vert^2 \vert \psi_z(z) \vert^2 \ ,  
\label{rho}
\end{eqnarray}
and using the Eqs.~(\ref{funcaox}),~(\ref{funcaoyy})~and~(\ref{funcaozz}), it yields:
\begin{eqnarray}
\rho_x(x)&=&\vert \psi_x(x) \vert^2 \nonumber \\ 
&=& \sum_{\mu\nu} N_\mu N_\nu C_\mu^\ast C_\nu (x-X_\mu)^{n_\mu}	(x-X_\nu)^{n_\nu} \times \nonumber \\
& & \exp\left[-\alpha_\mu(x-X_\mu)^2-\alpha_\nu(x-X_\mu)^2\right] \  \label{densidadex} \ ,\\
\rho_y(y) &=& \vert \psi_y(y) \vert^2 = N^2\exp(-2\alpha y^2) \label{densidadey} \ , \\
	\rho_z(z) &=& \vert \psi_z(z) \vert^2 = N^2\exp(-2\alpha z^2) \label{densidadez} \ .
\end{eqnarray}
Normalizing the densities $\rho_y(y)$ and $\rho_y(y)$ to unity we find $N^2 =\sqrt{2\alpha / \pi}$.

The probability density in momentum space is defined as
\begin{eqnarray}  
\gamma(p_x, p_y, p_z) &=& \gamma_x(p_x)\gamma_y(p_y)\gamma_z(p_z) \nonumber \\
&=& \vert \widetilde{\psi_x}(p_x) \vert^2 \vert \widetilde{\psi_y}(p_y) \vert^2 \vert \widetilde{\psi_z}(p_z) \vert^2  
\label{gamma}
\end{eqnarray}
where
\begin{eqnarray}
\gamma_x(p_x)&=&\vert \widetilde{\psi_x}(p_x) \vert^2 \nonumber \\
	&=&
	\dfrac{1}{2\pi}
	\sum_{\mu\nu} N_\mu N_\nu C_\mu^\ast C_\nu
		e^{ -\frac{p_x^2}{4}(1/\alpha_\mu+1/\alpha_\nu) +i{p_x(X_\mu-X_\nu)}}
	\sum_{k=0 \atop k\text{ even}}^{n_\mu}
	\sum_{\ell=0 \atop \ell\text{ even}}^{n_\nu}
	\binom{n_\mu}{k}
	\binom{n_\nu}{\ell}\times
	\nonumber \\ &&
	\Big(\dfrac{ip_x}{2\alpha_\mu}\Big)^{n_\mu-k}
	\Big(\dfrac{p_x}{2i\alpha_\nu}\Big)^{n_\nu-\ell}
	\dfrac{\Gamma\big(\dfrac{k+1}{2}\big)}{\alpha_\mu^{(k+1)/2}}
	\dfrac{\Gamma\big(\dfrac{\ell+1}{2}\big)}{\alpha_\nu^{(\ell+1)/2}} ,
	\label{densidadepx} \\ 
	\gamma_y(p_y) &=& \vert \widetilde{\psi_y}(p_y) \vert^2 = \dfrac{1}{\sqrt{2\alpha \pi}}\exp\left(-\dfrac{p_y^2}{2\alpha}\right) , \label{densidadepy} {\rm  \ and} \\
	\gamma_z(p_z) &=& \vert \widetilde{\psi_z}(p_z) \vert^2 = \dfrac{1}{\sqrt{2\alpha \pi}} \exp\left(-\dfrac{p_z^2}{2\alpha}\right)  . \label{densidadepz}
\end{eqnarray}
We present the details for determining of $\gamma_x(p_x)$ in the supplementary material.

\subsection{Shannon informational entropies}{\label{information_entropies}}

In the context of atomic and molecular physics, Shannon informational entropies in the space of positions, $S_r$, and momentum, $S_p$, can be written as~\cite{sen2011,nascimento-prudente2018}
\begin{eqnarray}
S_r=-\int \rho(x,y,z) \ln \left( \rho(x,y,z) \right) d^3r
\label{Sr}
\end{eqnarray}
and
\begin{eqnarray}
S_p=-\int\gamma(p_x,p_y,p_z) \ln \left( \gamma(p_x,p_y,p_z) \right) d^3p . 
\label{Sp}
\end{eqnarray}
The probability densities $\rho(x,y,z)$ and $\gamma(p_x,p_y,p_z)$ are defined as in Eqs.~(\ref{rho})~e~ (\ref{gamma}). Adopting $\rho(x,y,z)$ normalized to unity the $S_r$ entropy can be written as
\begin{eqnarray}
S_r&=& S_x +S_y +S_z \ ,
\label{Srp}
\end{eqnarray}
where
\begin{eqnarray}
S_x &=& -\int \rho_x(x) \ln \left( \rho_x(x) \right) dx , \label{Sx} \\ 
S_y &=& -\int \rho_y(y) \ln \left( \rho_y(y) \right) dy , \label{Sy} \textrm{\ and} \\
S_z &=& -\int \rho_z(z) \ln \left( \rho_z(z) \right) dz \label{Sz} .
\end{eqnarray}
Analogously, using $\gamma(p_x,p_y,p_z)$ normalized to unity the $S_p$ entropy becomes 
\begin{eqnarray}
S_p&=& S_{p_x} +S_{p_y} +S_{p_z}  ,
\label{Spp}
\end{eqnarray}
where
\begin{eqnarray}
S_{p_x} &=& -\int \gamma_x(p_x) \ln \left( \gamma_x(p_x) \right) dp_x \ , \label{Spx} \\ 
S_{p_y} &=& -\int \gamma_y(p_y) \ln \left( \gamma_y(p_y) \right) dp_y, \label{Spy} \textrm{\ and} \\
S_{p_z} &=& -\int \gamma_z(p_z) \ln \left( \gamma_z(p_z) \right) dp_z. \label{Spz}
\end{eqnarray}
The quantities $S_r$ and $S_p$ are interpreted as measures of delocalization or localization of the probability distribution~\cite{Corzo-et_al2012,cruz-et_al2021}.

We determine the entropies $S_y$ and $S_z$ analytically by replacing Eqs.~(\ref{densidadey})~e~(\ref{densidadez}) in Eqs.~(\ref{Sy})~and~~ (\ref{Sz}), respectively, so that,
\begin{eqnarray}
	S_y = S_z=-\dfrac{1}{2}\ln\left(\dfrac{2\alpha}{\pi}\right)+\dfrac{1}{2} .
\end{eqnarray}
Computing such results in Eq.~(\ref{Srp}) we have
\begin{eqnarray}
	S_r &=& S_x - \ln\left(\dfrac{2\alpha}{\pi}\right) + 1 \ .
	\label{Sr2}
\end{eqnarray}
 $S_x$ is calculated numerically using the density~(\ref{densidadex}) in Eq.~(\ref{Sx}).

Similarly, we obtain the values of $S_{p_y}$ and $S_{p_z}$ by substituting Eqs.~(\ref{densidadepy})~and~(\ref{densidadepz}) in Eqs.~(\ref {Spy})~and~~(\ref{Spz}), so that  
\begin{eqnarray}
	S_{p_y} = S_{p_z} = \frac{1}{2}+\frac{1}{2}\ln(2\pi\alpha)  .
\end{eqnarray}
Considering such results in Eq.~(\ref{Spp}) one gets
\begin{eqnarray}
	S_p= S_{p_x} + \ln(2\pi\alpha) + 1 .
	\label{Sp2}
\end{eqnarray}
$S_{p_x}$ is calculated numerically using the density~(\ref{densidadepx}) in Eq.~(\ref{Spx}).

The sum $S_t$ is composed of the addition of the quantities $S_r$ and $S_p$ which, in  turn, originate the entropic uncertainty principle mathematized as~\cite{birula-mycielski1975}
\begin{eqnarray}
S_t&=&S_r+S_p  \nonumber \\
&=& - \int  \rho(x,y,z)   \gamma(p_x,p_y,p_z)  \ln \left[ \rho(x,y,z)   \gamma(p_x,p_y,p_z)  \right] d^3rd^3p \nonumber  \\ 
&\geq & 3(1+\ln\pi) .
\label{St}
\end{eqnarray}
The value of $S_t$ is limited by the relation~(\ref{St}) which exhibits a minimum value. From the entropic uncertainty relation we can derive the Kennard uncertainty relation. More specifically, adding the
Eqs.~(\ref{Sr2})~e~(\ref{Sp2}) we obtain
\begin{eqnarray}
	S_t= S_x + S_{p_x} + 2 \left[ \ln (\pi) + 1 \right] \geq 3(1+\ln\pi) \ .
	\label{St2}
\end{eqnarray}
Note that this last expression does not depend on $\alpha$.

Shannon entropies are dimensionless quantities from the point of view of physics. However, subtleties surround this issue since, in principle, we have quantities that have physical dimensions in the argument of the logarithmic function. For a more detailed discussion on this topic see Refs.~\cite{nascimento-prudente2018,nascimento-et_al2020,matta-etal2011}.

\section{Analysis and Discussion}

In this Section, the energy and the entropic quantities $S_r$, $S_p$ and $S_t$ determined by Eqs.~(\ref{Sr2}),~(\ref{Sp2})~and~(\ref{St2}) are discussed as a function of the parameter $A_2$ and, afterwards, as function of $k$. In the first case we have kept fixed $A_1$, $k$ and $V_0$; and in the second case, $A_1$, $A_2$ and $V_0$. The specific/fixed values of the parameters in question, besides the values of $m_c$ and $V_0$ are based in Ref.~\cite{duque-2023-magnetico,duque2023-laser}.

The calculations in our study are performed in atomic units (a.u.). In order to compare some of our results with those previously published in the literature we adopt in this section the parameter $k$ in nanometer (nm) and, more specifically, we highlight the energetic contribution along the $x$-axis in meV. 

The optimized wave function was expanded into the following basis functions: on the $x$ axis we employ orbitals of the type 2s2p2d2f2g (in total 10 functions located in each well) and on the $y$ and $z$ axes, 1s type orbitals. In cases where states are degenerate, symmetrization and antisymmetrization were done.

\subsection{Energy analysis}{\label{energias}}

\begin{figure}[h]
\centering
%\hspace*{-1.2 cm}
\includegraphics[scale=0.4]{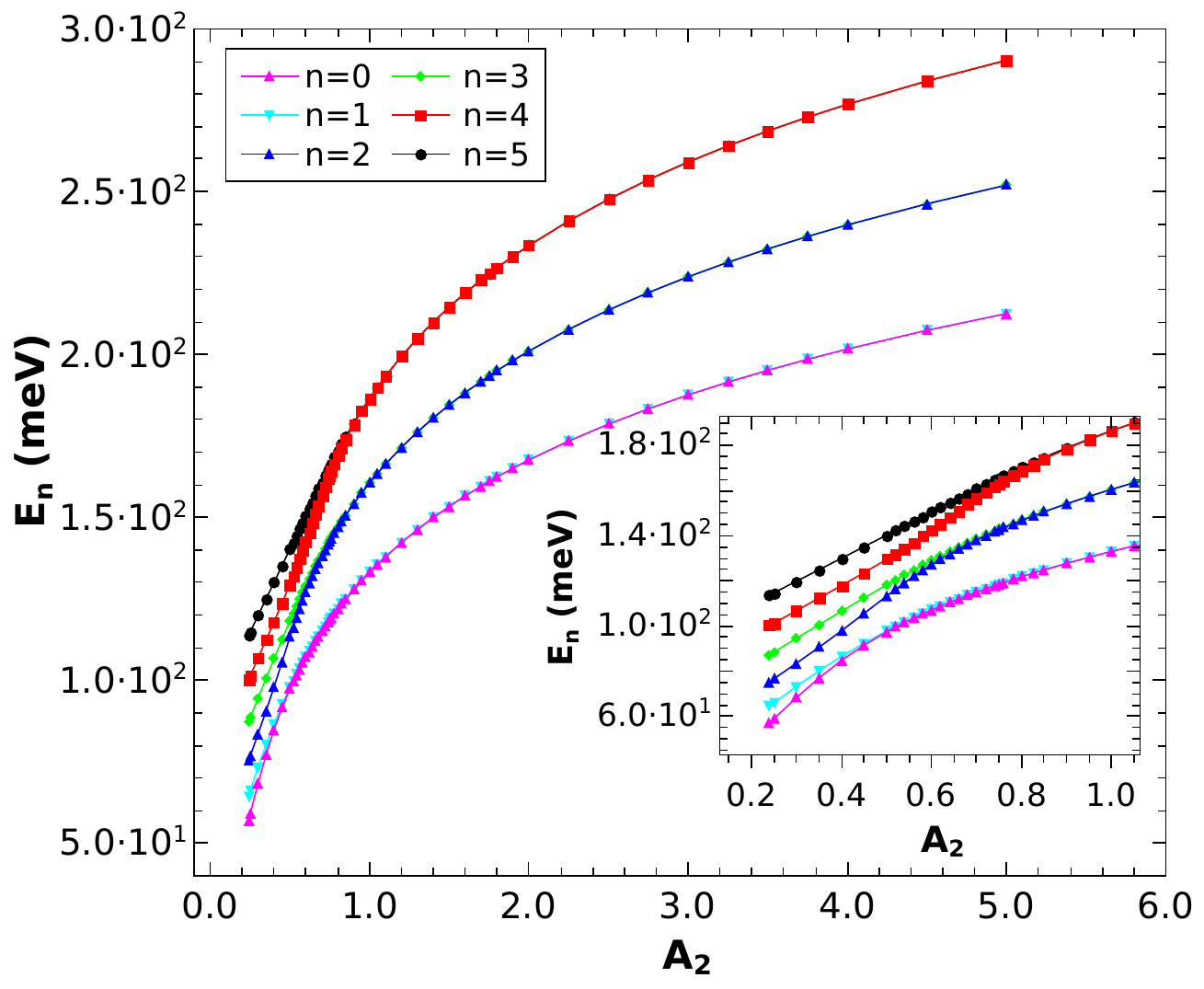}
%\vspace{-1.2 cm}
\caption{Energy contribution along the $x$-axis for states $n=0-5$ as a function of $A_2$, for $A_1=$ 0.200, $k=$ 20.000~nm  and  V$_0 =$ 228.00~meV. The inset details the region where the states are completely non-degenerate and merge two by two into one.}
\label{EXA3}
\end{figure}

We present in Fig.~\ref{EXA3} the energy curves $E_n$ (contribution along the $x$-axis) as a function of the parameter $A_2$ ranging from 0.240 to 5.000 for the first six quantum states $(n = 0 - 5)$ with $A_1$ = 0.200, k = 20.000 nm  and  V$_0$ = 228.00 meV. Inset: it is detailed the energy curves for the $A_2$ ranging from 0.240 to 1.050, where states initially non-degenerate become degenerate two by two as $A_2$ increases. In Table~S1 of the supplementary material you can find the energy values as a function of $A_2$.

According to the Table~S1, we have that the degeneracy for states $n$ = 0 and $n$ = 1 appears in the interval of $1.200 \leq A_2 \leq 1.400$, and at $A_2$ = 1.300 we have $E_0 = E_1 =$ 146.24146~meV. The degeneracies in $n$ = 2 and $n$ = 3 originate at values of $1.400 \leq A_2 \leq 1.600$, and at $A_2$ = 1.500 we find $E_2$ = $E_3$ = 184.54417~meV. Finally, the degeneracies in $n=4$ and $n$ = 5 begin between $1.800 \leq A_2 \leq 2.000$, and at $A_2$ = 1.900 we have $E_4$ = $E_5$ = 230.14020~meV. Otherwise, we observe by inset of Fig.~\ref{EXA3} that the decrease in the values of $A_2$ causes the system to rely on non-degenerate states. 

In Fig.~4a we present the probability density curves in the position, $\rho_x(x)$ , as a function of $x$ for the ground state and different values of $A_2$. In the curves of $\rho_x(x)$ for $A_2$ = 0.240 and 0.400 the state is not degenerate, in this case, the electron has the probability of being in one or both wells of the function $V_{DQD}(x)$, and even above the internal barrier. In the curves of $\rho_x(x)$ for $A_2$ = 1.300 and 1.400 the state is degenerate and the electron has the probability of being in only one of the wells of $V_{DQD}(x)$. For completeness, in Fig.~4b we present the probability density curves in the momentum, $\gamma_x(p_x)$, as a function of $p$ for the ground state for $A_2$ = 0.240, 0.400, 1.300 and 1.400.

\begin{figure}[h] % Inicia o ambiente de figuras
  \subfigure{ % Começa a incluir a figura fig1.pdf
    \includegraphics[scale=0.36]{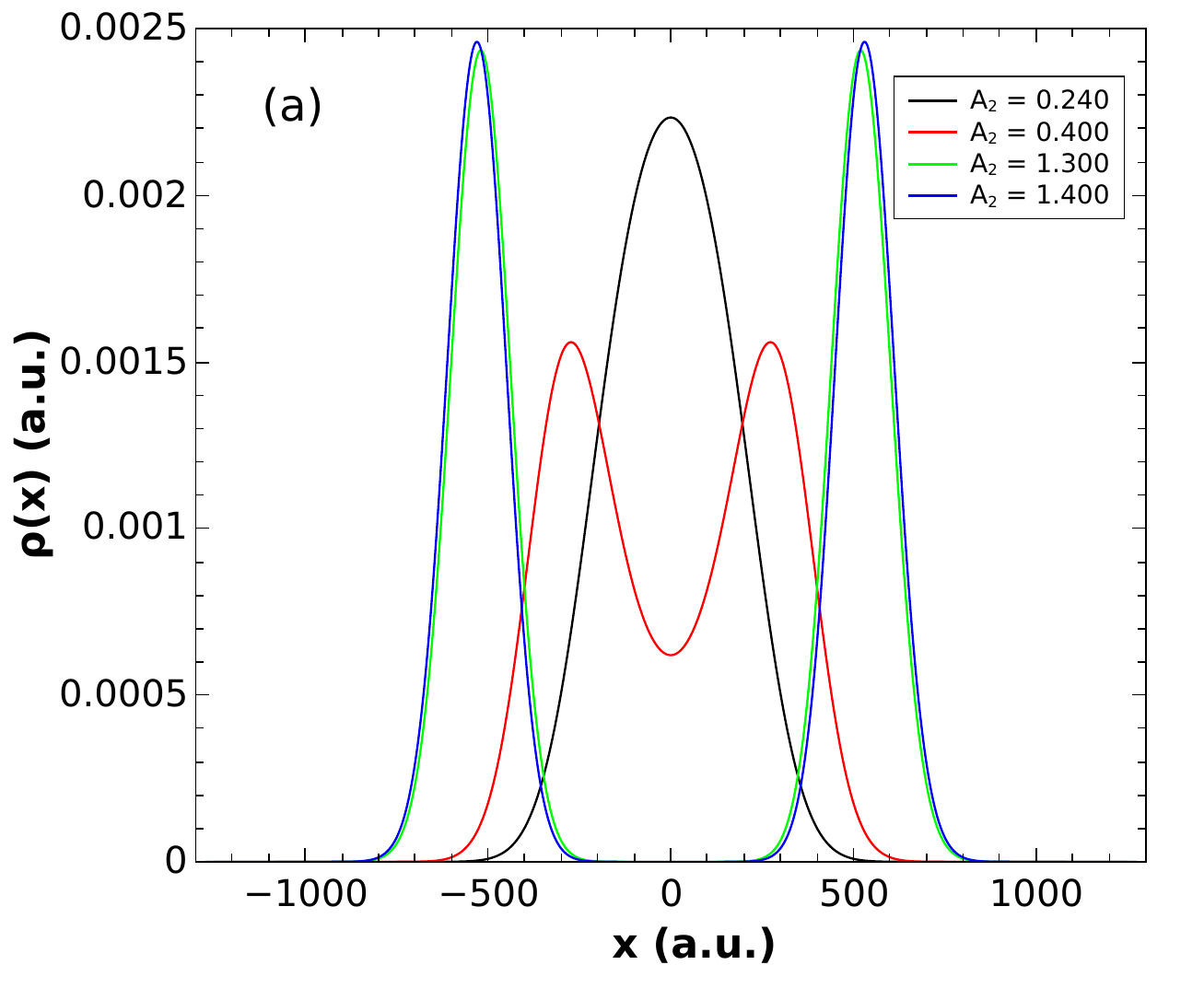}
  } % Termina de incluir a figura fig1.pdf
  \subfigure{ % Começa a incluir a figura fig2.pdf na mesma linha da figura fig1.pdf
    \includegraphics[scale=0.35]{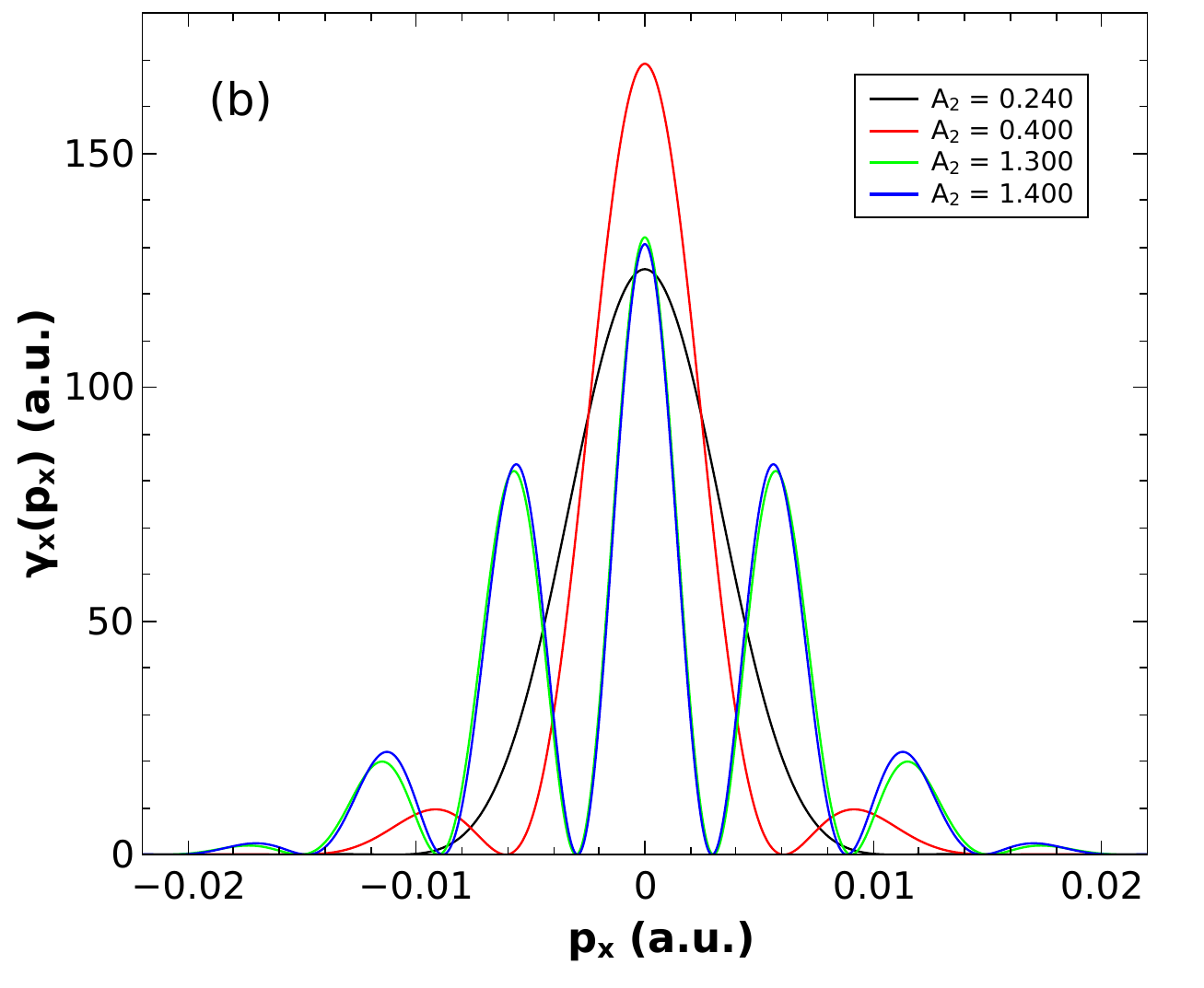}
  } % Termina de incluir a figura fig2.pdf
\caption{Probability densities $\rho_x(x)$ and $\gamma_x(p_x)$ in the position and momentum space, respectively, for the ground state with different values of $A_2$, for $A_1$ = 0.200, $k$ = 20.000~nm and  V$_0 =$ 228.00~meV.} 
\label{densidadesa2}
\end{figure} % Fecha o ambiente de figuras

\begin{figure}[h]
\centering
%\hspace*{-1.2 cm}
\includegraphics[scale=0.4]{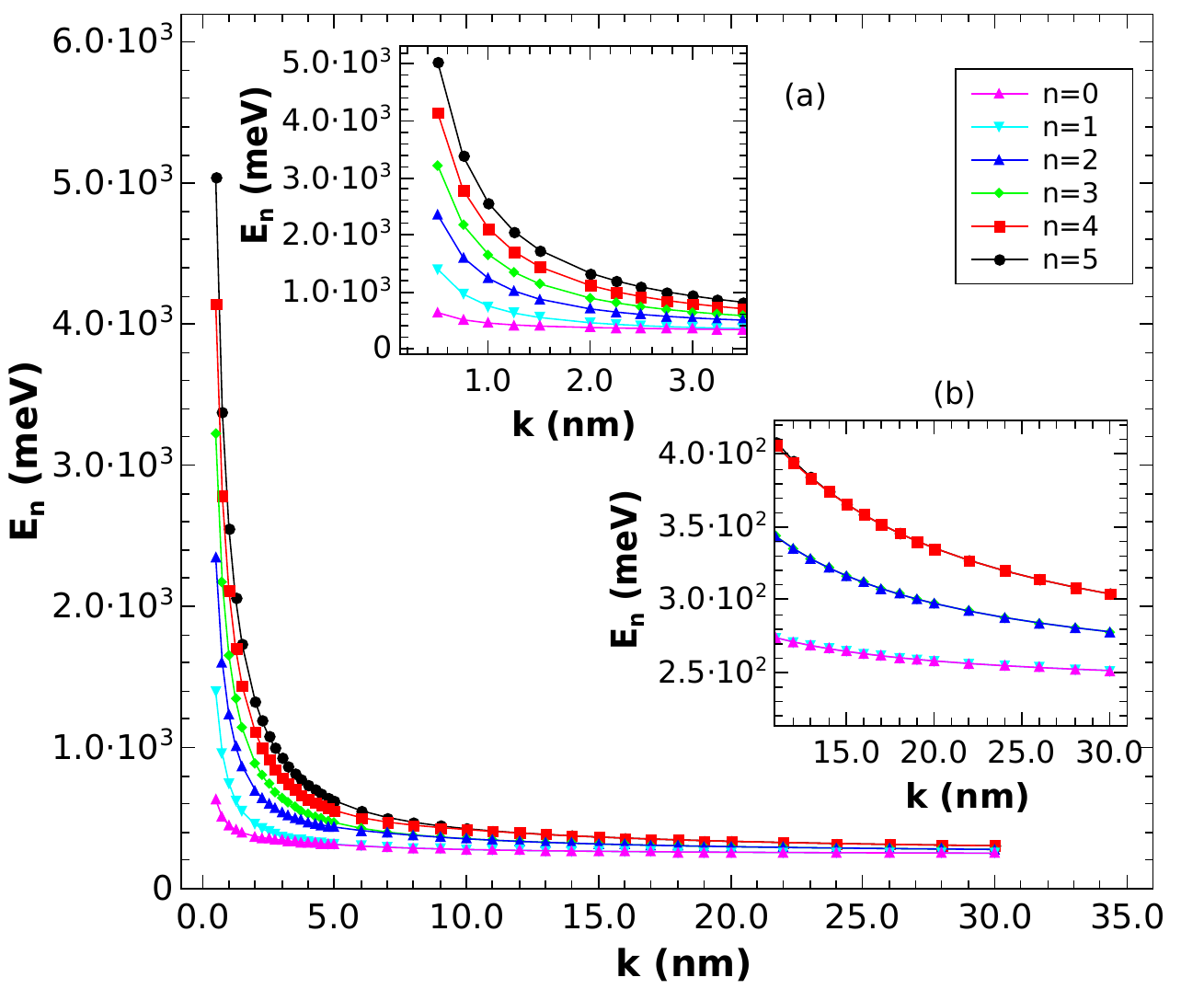}
%\vspace{-1.2 cm}
\caption{Energy contribution along the $x$-axis for states $n$ = 0-5 as a function of $k$, for $A_1$ = 0.400,  $A_2$ = 2.000  and  V$_0$ = 228.00 meV. The insets show the non-degenerate region of the energy curves (0.5 $\leq$ k $\leq$ 3.5 nm) and the degenerate one (11.0 $\le$ k $\leq$ 30.0 nm).}
\label{EXk}
\end{figure}

In the main graph of Fig.~\ref{EXk} we present the energy curves $E_n$ (contribution along the $x$-axis) for the first six quantum states $(n=0-5)$ as a function of the parameter $k$ varying from 0.500~nm to 30.000~nm with $A_1$ = 0.400, $A_2$ = 2.000 and V$_0$ = 228.00 meV. Furthermore, we indicate in the insets (a) the energy curves with the parameter $k$ varying from 0.500~nm to 3.500~nm and (b) the energy curves with $k$ varying from 11.000~nm to 30.000~nm. In Table~S2 of the supplementary material the values obtained for energies as a function of $k$ can be found.

We observe in the main graph of Fig.~\ref{EXk} and inset (b) that with the increase in the values of $k$ the energies merge two by two into one, that is, $E_0=E_1$, $E_2=E_3$, $E_4=E_5$. By inset (a), with the decrease in the values of $k$ and the increase in the effects of confinement, we identify the appearance of non-degenerate states, besides, we have on considerable increase in the values of $E_n$. 

\begin{figure}[h] % Inicia o ambiente de figuras
  \subfigure{ % Começa a incluir a figura fig1.pdf
    \includegraphics[scale=0.36]{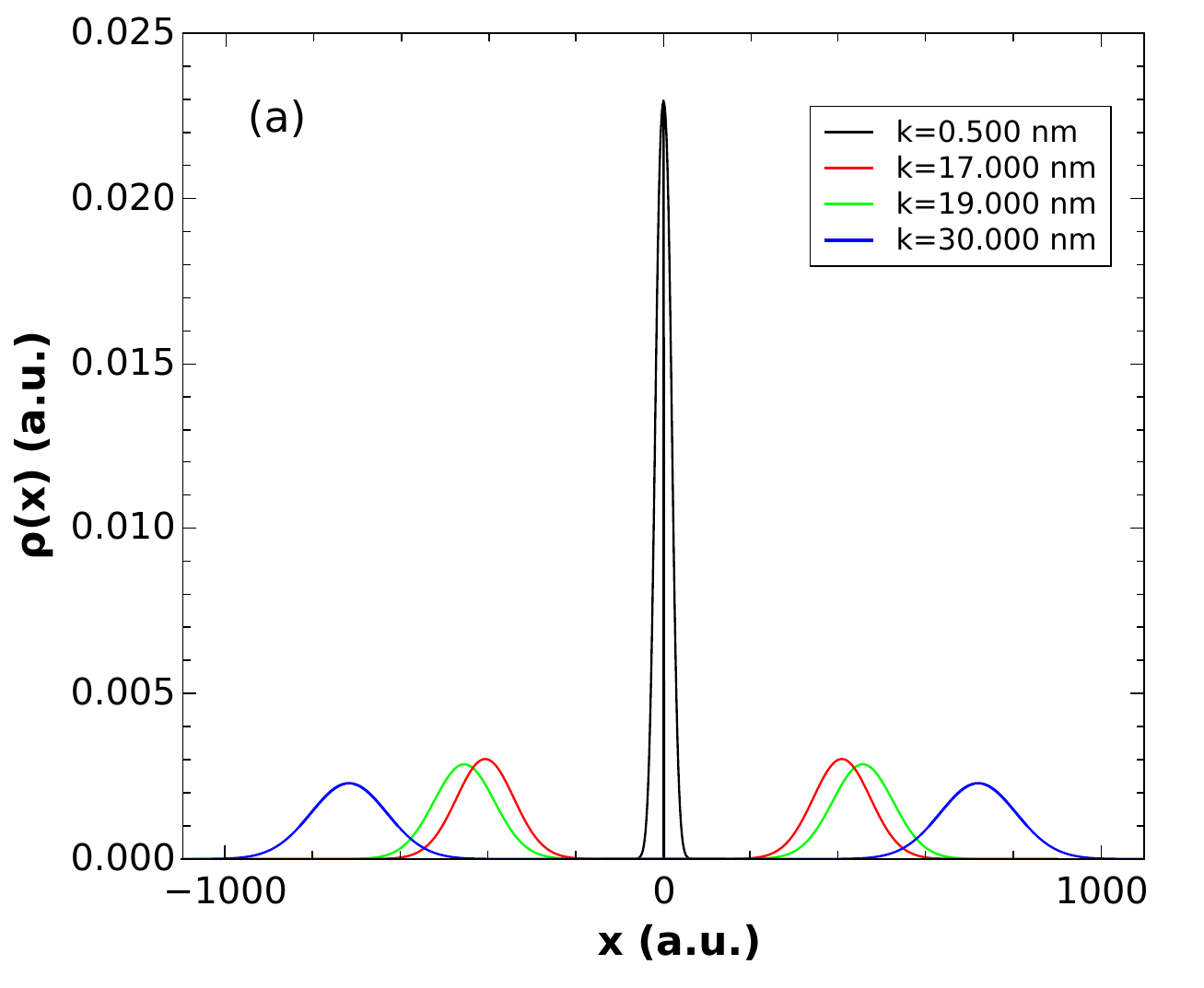}
  } % Termina de incluir a figura fig1.pdf
  \subfigure{ % Começa a incluir a figura fig2.pdf na mesma linha da figura fig1.pdf
    \includegraphics[scale=0.35]{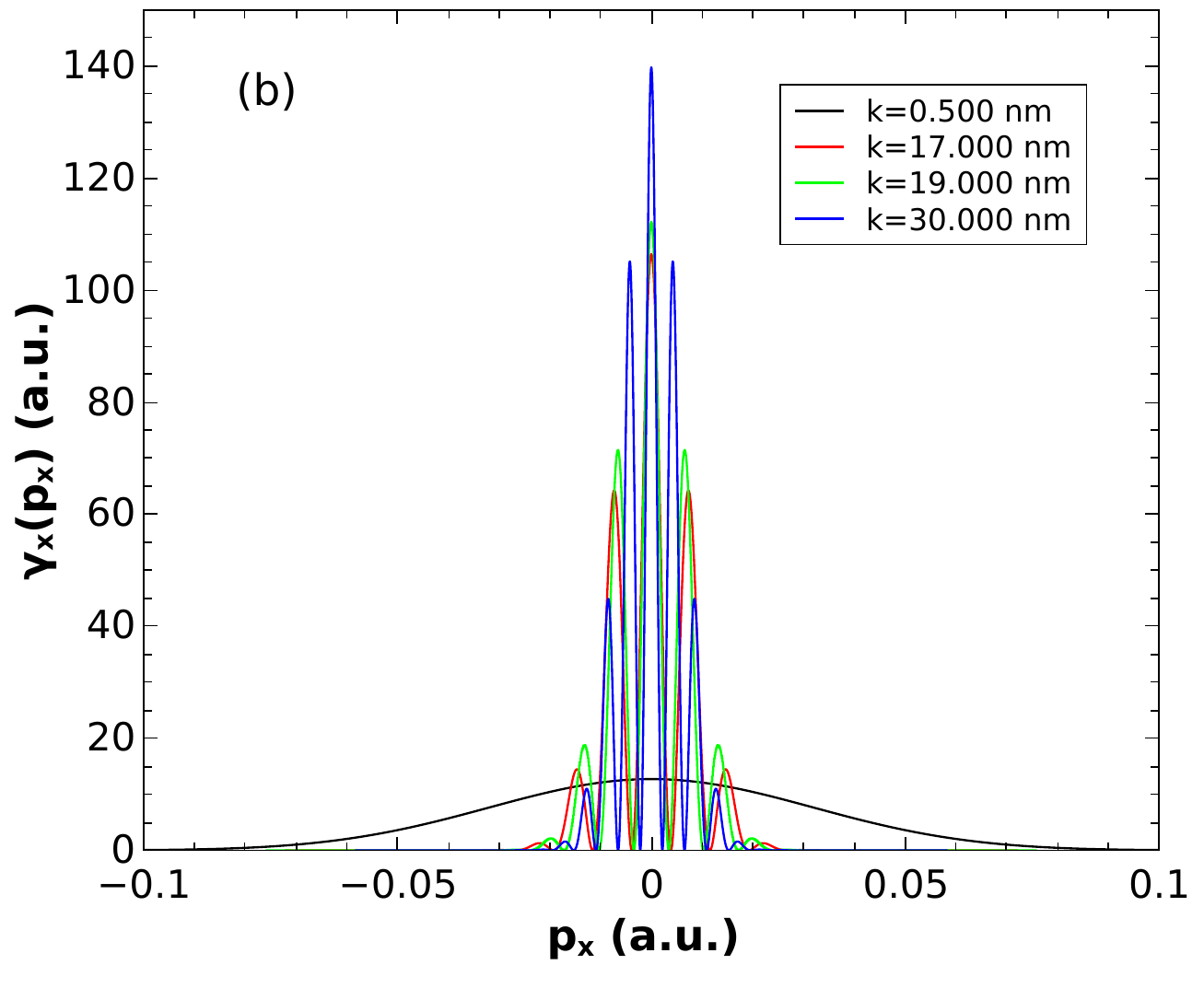}
  } % Termina de incluir a figura fig2.pdf
\caption{Probability densities $\rho_x(x)$ and $\gamma_x(p_x)$ in the position and momentum space, respectively, for the ground state with different values of $k$, for $A_1$ = 0.400, $A_2$ = 2.000 and V$_0$ = 228.00 meV.}
\label{densidadesk2}
\end{figure}

In the graphs of Fig.~\ref{densidadesk2} we present the probability density curves for the ground state in the position and momentum space, $\rho_x(x)$ (Fig.~~\ref{densidadesk2}a) and $\gamma_x(p_x)$ (Fig.~\ref{densidadesk2}b), respectively, for different values of $k$. In both cases, $\rho_x(x)$ and $\gamma_x(p_x)$, for the value of $k$ = 0.500 nm the ground state is non-degenerate, otherwise, for $k$ = 17.000 nm, 19.000 nm and 30.000 nm the state is degenerate. In this way, we perceive changes in the shape of the probability distributions  when the values of $k$ imply or not degeneracy in energies.

 % Fecha o ambiente de figuras

As long as comparison has been possible, we have obtained a good agreement with the values found in Ref.~\cite{duque-2023-magnetico,duque2023-laser} regarding the energy as a function of $A_2$ and $k$. As in the present work, Duque \textit{et al} have also used matrix diagonalization methods; however, they have adopted an expanded wave function in terms of orthonormal trigonometric functions.

\subsection{Informational Analysis}

\begin{figure}[h]
\centering
%\hspace*{-1.2 cm}
\includegraphics[scale=0.4]{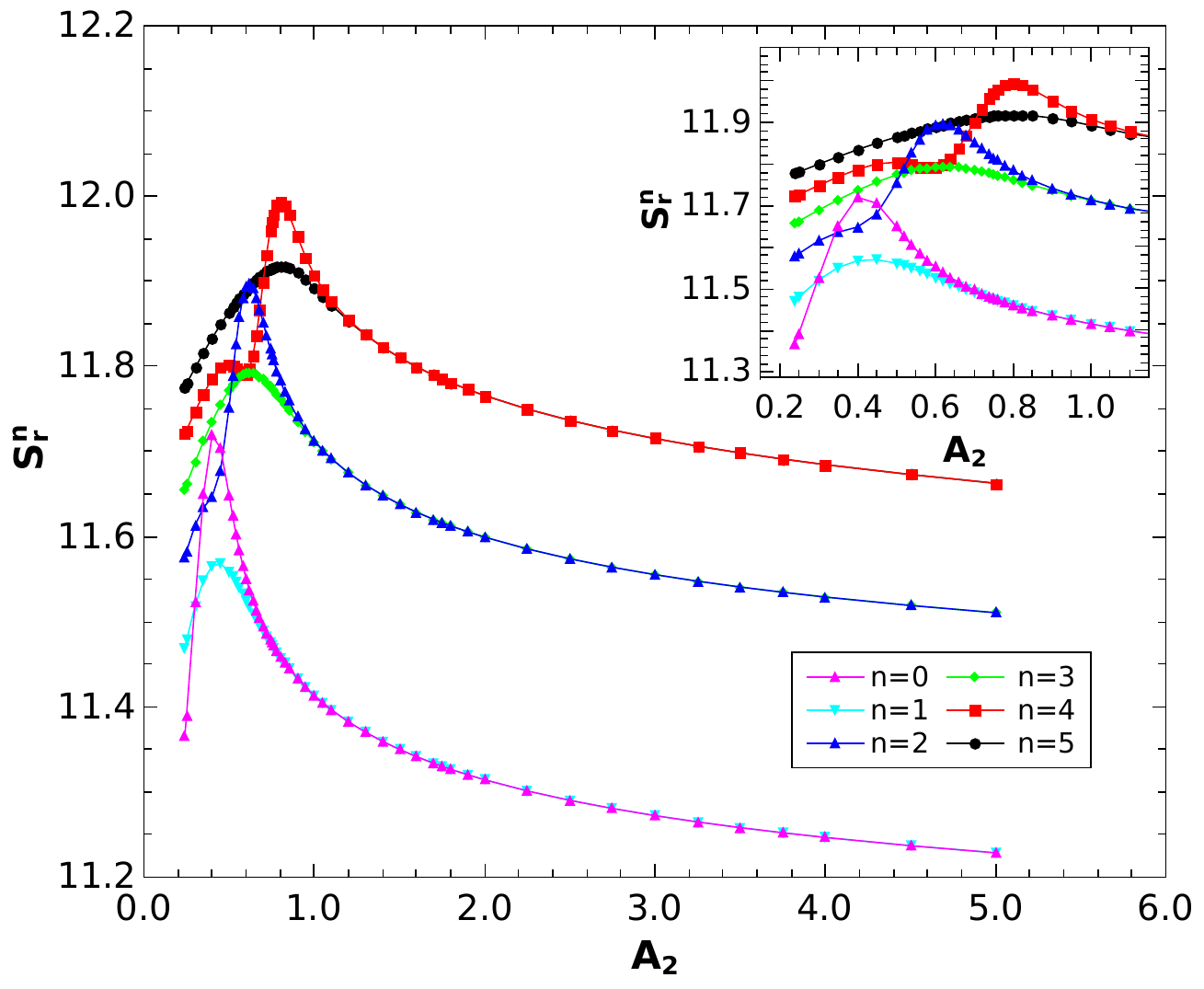}
%\vspace{-1.2 cm}
\caption{Shannon entropy, $S^n_r$, for states $n=0-5$ as a function of $A_2$ ranging from 0.240 to 5.000, for $A_1$ = 0.200,  $k$ = 20.000 nm  and  V$_0$~=228.00 meV. In the inset, it is detailed the initial behaviour of $S^n_r$ as a function of $A_2$.} 
\label{SrXA2}
\end{figure}

In the main graph of Fig.~\ref{SrXA2} we present the curves of Shannon entropy  in the space of positions for the first six quantum states $(n=0-5)$, $S^n_r$, as a function of the parameter $A_2 $ ranging from 0.240 to 5.000 with $A_1$ = 0.200,  $k$ = 20.000 nm  and  V$_0$~=228.00 meV.  The inset details the curves of $S^n_r$ in the range of $A_2=$0.240 to 1.100. In Table~S3 of the supplementary material we provide the values of $S^n_r$ as a function of $A_2$.

According to Fig.~\ref{SrXA2} we notice that the values of the Shannon entropy curves $S^n_r$ get closer two by two as $A_2$ increases, i.e., $S^0_r \rightarrow S^1_r$, $S^2_r \rightarrow S^3_r$ and $S^4_r \rightarrow S^5_r$. In Table~S3, we identified that the degeneracy of $S^n_r$ for states $n=0$ and $n=1$ appears in the values of $A_2$ comprised between $1.300 \leq A_2 \leq 1.500$, and at $A_2=$ 1.400 we have $S^0_r=S^1_r=$ 11.35961. The degeneracy in $S^n_r$ for states $n=2$ and $n=3$ becomes evident in the interval $1.400 \leq A_2 \leq 1.600$, and at $A_2=$ 1.500 we find $S^2_r=S ^3_r=$ 11.63816. Finally, although at $A_2=$ 1.750 we already have $S^4_r=S^5_r=$ 11.78527, the degeneracy in $S^n_r$ for $n=4$ and $n=5$ appears in the values between $1.800 \leq A_2 \leq 2.000$, and at $A_2=$ 1.900 we have the identical value of $S^4_r=S^5_r=$ 11.77304.

In Table~\ref{regiaodegenerescencia} we present the regions of $A_2$ where the degeneracies in energy and entropy $S^n_r$ for states $n=0-5$ originate. For states $n=0$ and $n=1$, the range of values of $A_2$ where this occurs coincides reasonably with the range of values of $A_2$ where the degeneracy in $S^n_r$ originates. For states $n=2$ and $n=3$, and also for $n=4$ and $n=5$, the set of values of $A_2$, where the degeneracy in energies and in $S_r$ originate, are identical. In this way, we conjecture that the Shannon entropy in the space of positions, $S^n_r$, can successfully map the degeneracy of states when we vary the values of $A_2$ in the double quantum dot studied.

\begin{table}[h]
%\centering
\caption{Range  of $A_2$ where degeneracies in energies and entropy $S^n_r$ for states $n=0-5$  originate.}
\begin{tabular}{|c|cc|}
 \hline
          & \multicolumn{2}{c|}{Range  of $A_2$}                                                     \\ \hline
States   & \multicolumn{1}{c|}{Energy}                                  & $S_r$                                   \\ \hline
n=0 e n=1 & \multicolumn{1}{c|}{1.200 \textless{} $A_2$ \textless 1.400}   & 1.300 \textless{} $A_2$ \textless 1.500 \\ \hline
n=2 e n=3 & \multicolumn{1}{c|}{1.400 \textless{} $A_2$ \textless{}1.600} & 1.400 \textless{} $A_2$ \textless 1.600 \\ \hline
n=4 e n=5 & \multicolumn{1}{c|}{1.800 \textless{} $A_2$ \textless 2.000}  & 1.800 \textless{} $A_2$ \textless{}2.000 \\ \hline
\end{tabular}
%\vspace{0.5 cm}
\label{regiaodegenerescencia}
\end{table}

By analyzing the inset of Fig.~\ref{SrXA2} we see that as $A_2$ decreases, the values of $S^n_r$ increase until they reach a maximum value and, from that point on, they decrease again. The oscillations for the values of $S^n_r$ can be justified taking into account that the information entropies reflect a measure of the delocalization/localization  of $\rho_x(x)$. For example, for the ground state, according to Table~S3, we highlight that: at $A_2=1.300$ we have $S^0_r=$ 11.37033, at $A_2=$ 0.400 we have a maximum value of $S^0_r=$ 11.71876 and, finally, at $A_2=$ 0.240 we find $S^0_r=$ 11.36528. These values of $S^n_r$ agree with Fig.~\ref{densidadesa2}(a), since the delocalization of the green curve is smaller than that of the red curve which, in turn, is greater than that of the black curve. Similar analyzes can be undertaken for other states. 

Taking the state $n=0$ in Fig.~\ref{SrXA2}, starting from $A_2=0.240$, where $S^0_r=11.36528$, the values of $S^0_r$ increase up to the maximum value of $S^{ 0}_r=11.71876$ in $A_2=0.400$. From this point of maximum entropy, the values of $S^0_r$ decrease, passing through $A_2=1.300$ where $S_r=11.37033$. We observe in Fig.~\ref{potencialx}(a) that it is precisely at $A_2=0.400$ that the internal barrier of $V_{DQD}(x)$ begins to influence the decoupling between the two wells. Still, from Fig.~\ref{potencialx}(a), in $A_2=1.300$ the internal barrier of the function $V_{DQD}(x)$ is quite consolidated, strongly favoring the decoupling between the two wells of the function. In this way, we conjecture that the information entropy by means of $S^{0}_r$ is an indicator of the level of decoupling/coupling of the double quantum dot studied.

\begin{figure}[h]
\centering
%\hspace*{-1.2 cm}
\includegraphics[scale=0.4]{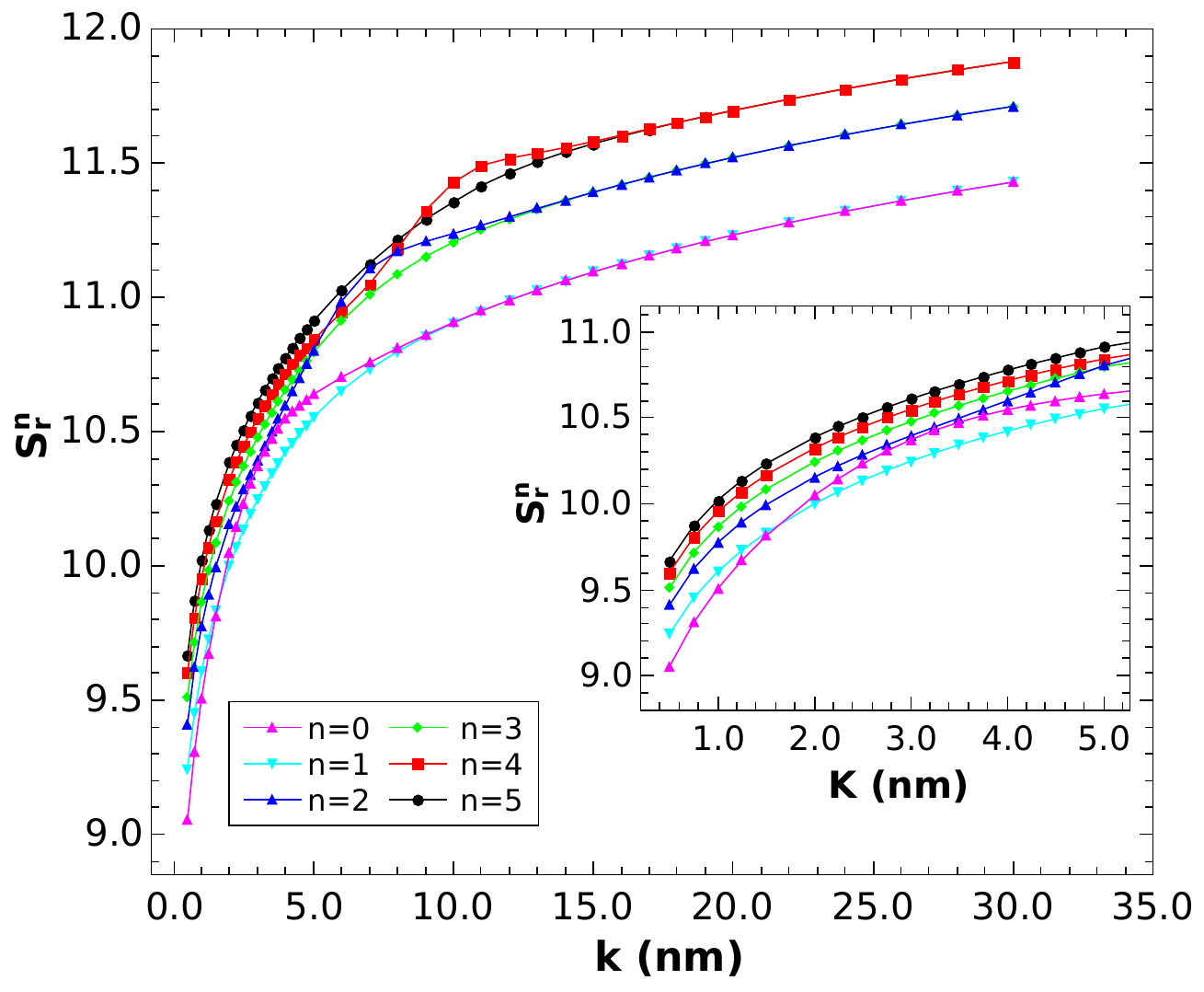}
%\vspace{-1.2 cm}
\caption{Shannon entropy, $S^n_r$, for states $n=0-5$ as a function of $k$ ranging from 0.500 to 30.000 nm, for $A_1$ = 0.400,  $A_2$ = 2.000  and  $V_0=228.00$~meV. In the inset, $S^n_r$ as a function of $k$ ranging from 0.500 to 5.000 nm.
} 
\label{SrXk}
\end{figure}

In the graph in Fig.~\ref{SrXk} we present for the first six quantum states $(n=0-5)$ the Shannon entropy curves in the space of positions, $S^n_r$, as a function of the parameter $k$ varying from 0.500~nm to 30.000~nm with $A_1 = 0.400$,  $A_2 = 2.000$  and  $V_0 = 228.00$~meV. In the inset we highlight $S^n_r$ in the range from $k=0.500$~nm to 5.000~nm. In Table~S4 of the supplementary material we provide the values of $S^n_r$ as a function of $k$.

According to the main graph of the Fig.~\ref{SrXk} when $k$ tends to infinity degeneracy arises in the values of $S_r$ such that $S^0_r=S^1_r$, $S^2_r=S^3_r$ and $S^4_r=S^5_r$. An increase in the values of $k$ widens the barriers of the potential function $V_{DQD}(x)$ reducing the effects of confinement. In this situation, the uncertainty in determining the location of the electron increases and, consequently, we identify an increase in the values of $S^n_r$. In fact, as $k$ increases the delocalization in $\rho_x(x)$ increases according to Fig.~\ref{densidadesk2} (a).

On the other hand, we observe in the inset of Fig.~\ref{SrXk} that as $k$ decreases there is a break  in the degeneracy of $S^n_r$. Furthermore, the decrease in $k$ generates a narrowing in the barriers of the potential function $V_{DQD}(x)$. In this case, there is an increase in the confinement situation and, consequently, a decrease in uncertainty in the location of the electron, causing $S^n_r$ values to decrease. Here, the delocalization in $\rho_x(x)$ decreases according to Fig.~\ref{densidadesk2}(a).

\begin{figure}[h]
\centering
%\hspace*{-1.2 cm}
\includegraphics[scale=0.4]{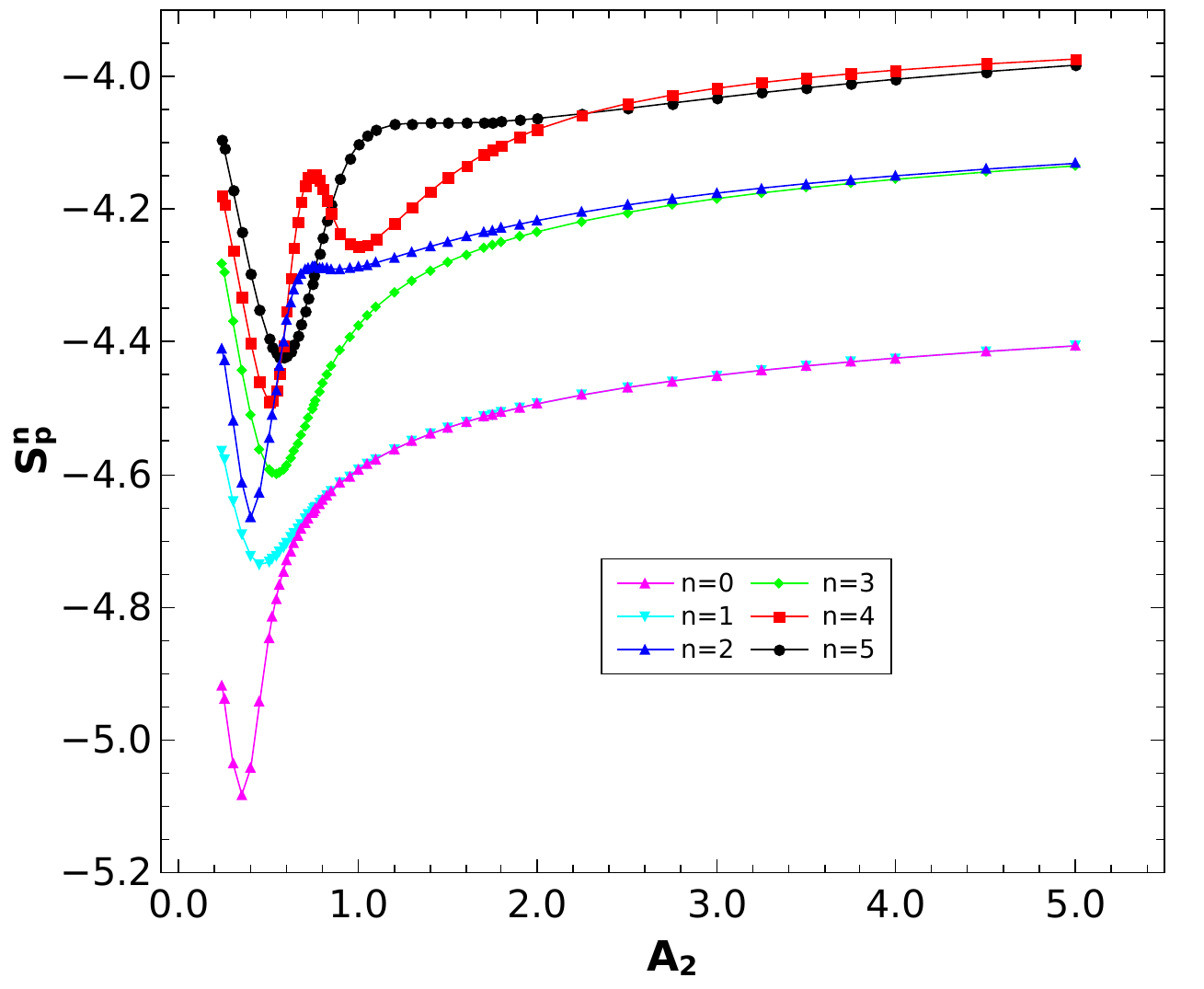}
%\vspace{-1.2 cm}
\caption{Shannon entropy, $S^n_p$, for states $n=0-5$ as a function of $A_2$ ranging from 0.240 to 5.000, for $A_1$ = 0.200,  $k$ = 20.000 nm  and  V$_0$ = 228.00 meV.}
\label{SpXA2}
\end{figure}

In Fig.~\ref{SpXA2} we display for the first six quantum states ($n=0-5$) the Shannon entropy curves in the momentum space, $S^n_p$, as a function of the parameter $A_2$ varying from 0.240 to 5.000 with $A_1$ = 0.200, $k$ = 20.000 nm and V$_0$ = 228.00 meV. In Table~S5 of the supplementary material we provide the values of $S^n_p$ as a function of $A_2$.

We have from Fig.~\ref{SpXA2} and according to Table~S5 that when $A_2$ increases occurs that $S^0_p \rightarrow S^1_p$. We did not identify degeneration in $S^{2-5}_p$. In general, with the decrease in $A_2$ the values of $S^n_p$ also decrease until they reach a minimum value, then increase again. Similar to what we did for $S^n_r$ the oscillatory behavior of the values of $S^n_p$ can be explained based in the curves of $\gamma_x(p_x)$ in the Fig.~\ref{densidadesa2}. 

\begin{figure}[h]
\centering
%\hspace*{-1.2 cm}
\includegraphics[scale=0.4]{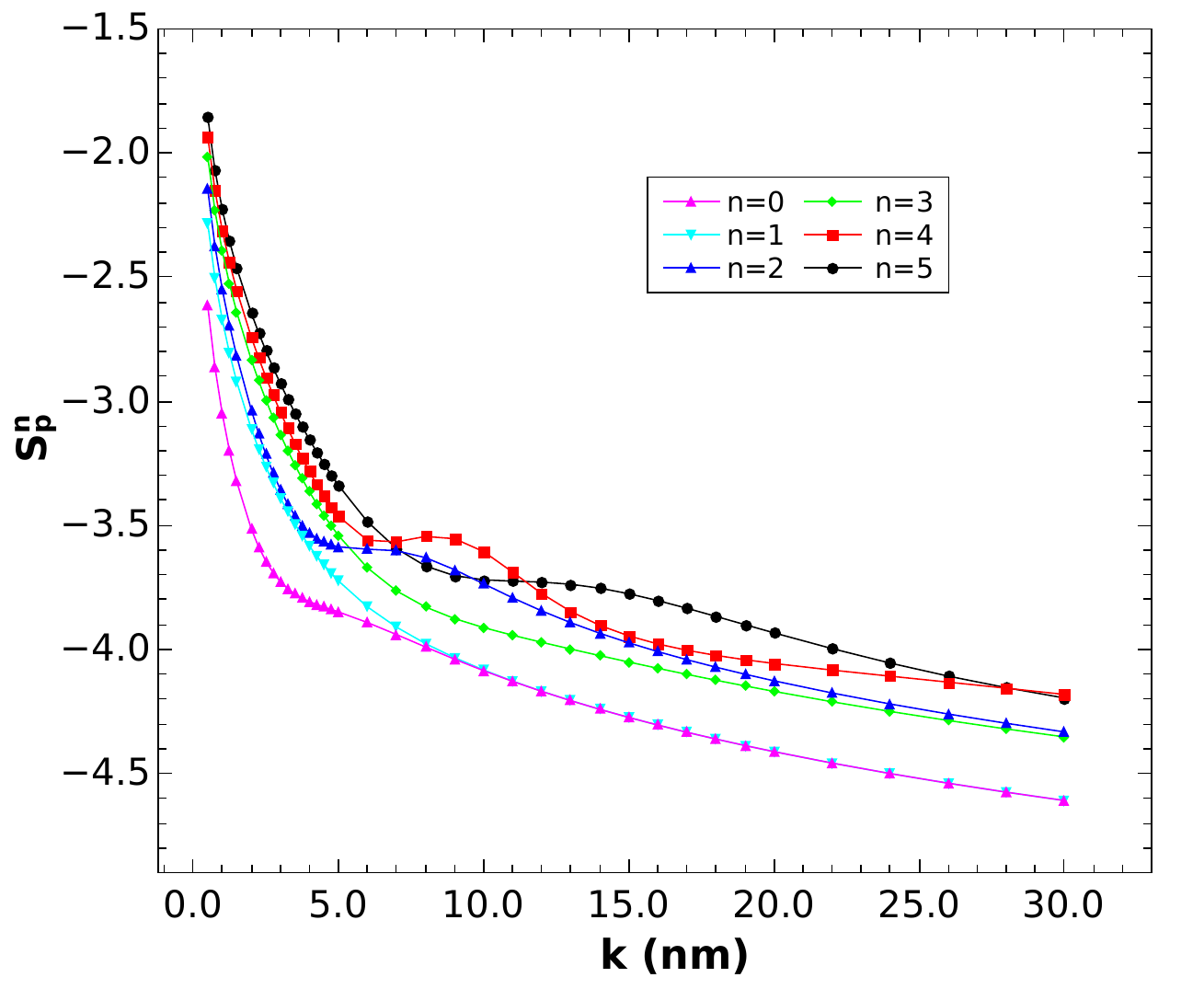}
%\vspace{-1.2 cm}
\caption{Shannon entropy, $S^n_p$, for states $n=0-5$ as a function of $k$ ranging from 0.500 to 30.000~nm, for $A_1$ = 0.400, $A_2$ = 2.000 and $V_0$ = 228.00 meV.} 
\label{SpXk}
\end{figure}

In Fig.~\ref{SpXk} we present for the first six quantum states $(n=0-5)$ the Shannon entropy curves in momentum space, $S^n_p$, as a function of the parameter $k$ varying from 0.500~ nm up to 30.000~nm with $A_1$ = 0.400, $A_2$ = 2.000 and $V_0$ = 228.00 meV. In Table~S6 of the supplementary material we have 
the values of $S^n_p$ as a function of $k$. 

According to the Fig.~\ref{SpXk} and Table~S6 when the values of $k$ increase we have degeneracy in $S^0_p$ and $S^1_p$. We did not identify degeneration in $S^{2-5}_p$. On the other hand, when $k$ decrease the values of $S^n_p$ increases. The behavior of the curve for $n$~=~4 shows intriguing oscillations. 

All values of $S^n_p$ obtained in this work are negative as can be seen in Tables~S5~e~S6 of the supplementary material. This result has an explanation in the quantum context~\cite{AQUINO20132062}, that is, when the limits of confinement are very small, the probability density becomes large and $\rho (x,y,z) > 1$. In this situation, $-\rho(x,y,z) \ln (\rho(x,y,z)) < 0$ and so $S_p$ (or $S_r$) can be negative. The original work by Shannon~\cite{shannon1948a} also indicates the possibility of obtaining negative values for informational entropy when working with continuous distributions.

\begin{figure}[h]
\centering
%\hspace*{-1.2 cm}
\includegraphics[scale=0.38]{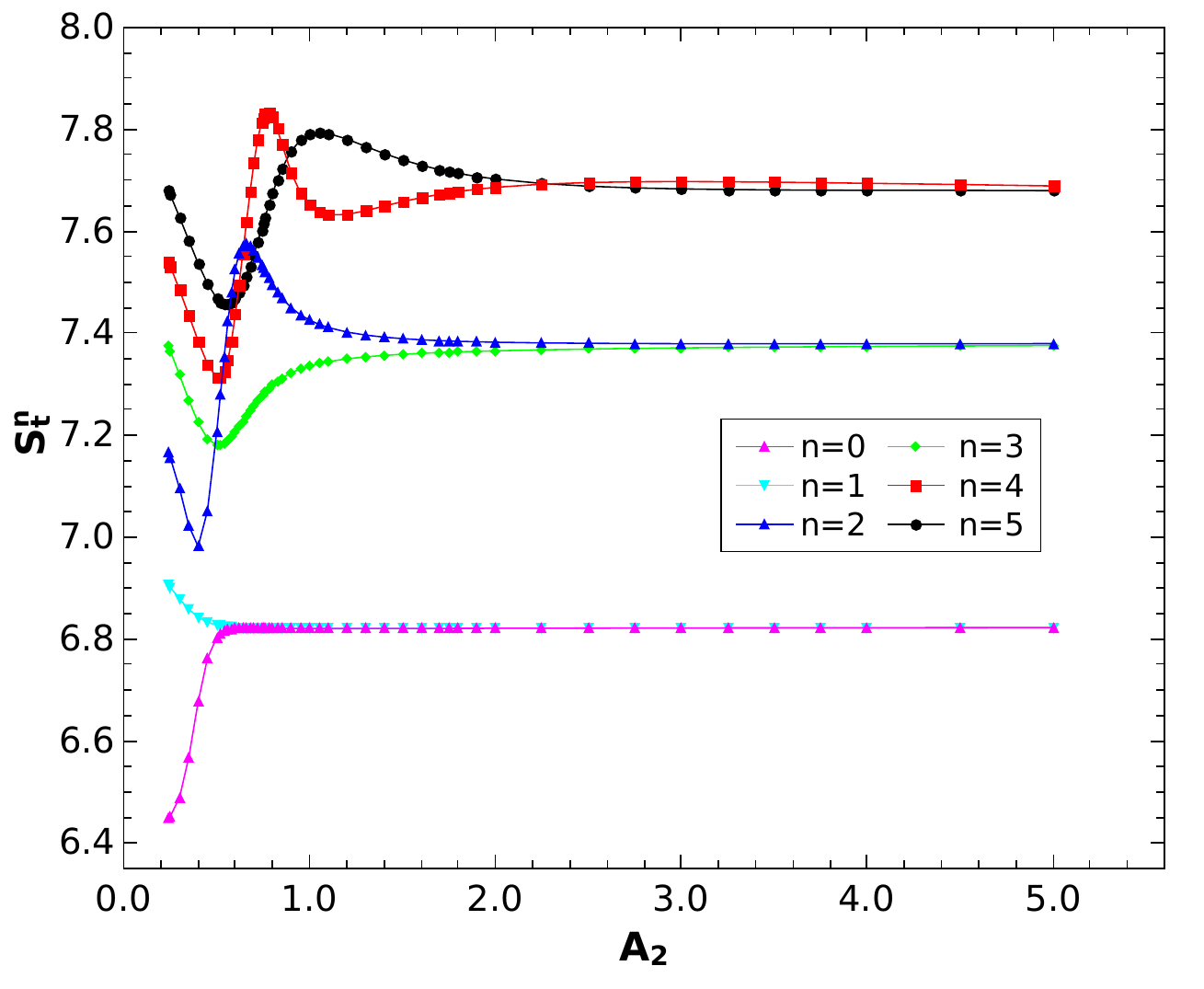}
%\vspace{-1.2 cm}
\caption{Entropy sum, $S^n_t$, for states $n=0-5$ as a function of $A_2$ ranging from 0.240 to 5.000, for $A_1$ = 0.200,  $k$ = 20.000 nm and V$_0$ = 228.00 meV.}
\label{StXA2}
\end{figure}

In Fig.~\ref{StXA2} we present for the first six quantum states $(n=0-5)$ the entropy sum curves $S^n_t$ as a function of the parameter $A_2$ varying from 0.240 to 5.000 with $A_1$ = 0.200,  $k$ = 20.000 nm and V$_0$ = 228.00 meV . In Table~S7 of the supplementary material we provide the values of $S^n_t$ as a function of $A_2$. 

From Fig.~\ref{StXA2} and Table~S7, as $A_2$ increases we have $S^0_t \rightarrow S^1_t$. We did not identify degeneracy in $S^{2-5}_t$. Furthermore, when $A_2$ tends to infinity $S^n_t$ tends to constant values. More specifically, the values of $S^0_t=6.44814$ for $A_2=0.240$ and $A_1=0.200$ (see green curve in Fig.~\ref{potencialx}a) is a value approximately equal to three times the value of the entropy sum for the one-dimensional harmonic oscillator in the ground state presented in Ref.~\cite{nascimento-et_al2020}.

We identify in Fig.~\ref{StXA2} oscillations in the $S^n_t$ curves with the occurrence of maximum and minimum values for states $n=2-5$. An elegant explanation for the extreme values of the entropy sum is presented in Ref.~\cite{liu2023}, that is, in general, the derivative of $S_t$ with respect to $A_2$ is given by $\frac{\partial S_t}{\partial A_2} = \frac{\partial S_r}{\partial A_2} + \frac{\partial S_p}{\partial A_2}$. Since the extreme points in the curves occur when $\frac{\partial S_t}{\partial A_2}=0$ with the absolute values $\vert \frac{\partial S_r}{\partial A_2} \vert$ and $ \vert \frac{\partial S_p}{\partial A_2} \vert$ being equal.

\begin{figure}[h]
\centering
%\hspace*{-1.2 cm}
\includegraphics[scale=0.38]{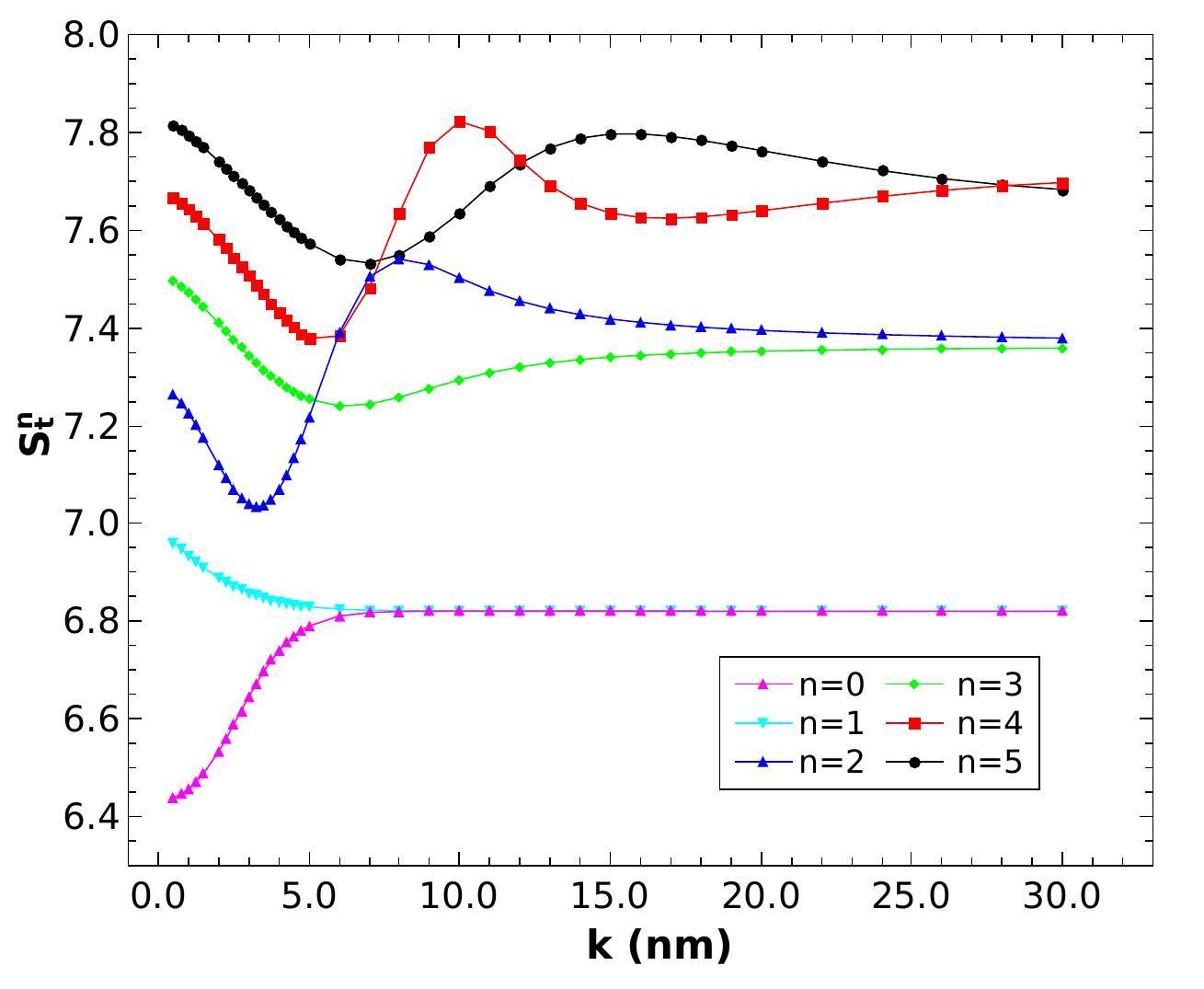}
%\vspace{-1.2 cm}
\caption{Entropy sum, $S^n_t$, for states $n=0-5$ as a function of $k$ ranging from 0.500 to 30.000 nm,  for $A_1$ = 0.400,  $A_2$ = 2.000  and  $V_0$ = 228.00 meV.}
\label{StXk}
\end{figure}

The Fig.~\ref{StXk} displays for the first six quantum states $(n=0-5)$ the entropy sum curves $S^n_t$ as a function of the parameter $k$ ranging from 0.500~nm to 30.000~nm with $A_1$ = 0.400,  $A_2$ = 2.000  and  $V_0$ = 228.00 meV. In Table~S8 of the supplementary material we provide the values of $S^n_t$ as a function of $k$.

As shown in Fig.~\ref{StXk} as $k$ grows, we see a degeneracy of $S^0_t$ and $S^1_t$. For $S^{2-5}_t$ we did not identify degeneration. When $k$ tends to infinity, the values of $S^n_t$ tend to constant values. With the decrease in the values of $k$, oscillations appear in the curves of $S^{2-5}_t$, such oscillations can be analyzed in a similar way as we presented previously, that is, taking into account the derivative of $S_t$ in relation to $k$.

The values of $S_r$ and $S_p$, whether as a function of $A_2$ or $k$, are compatible with the entropic uncertainty relationship. Furthermore, all values of $S_t$ obtained in this work (see Tables~S7~ and~S8) are above the minimum value of 6.43419 that the entropy sum can assume according to the Eq.~(\ref{St2}).

\section{Conclusion}

We have studied the electronic confinement in a double quantum dot using Shannon informational entropies. The confinement potential $\hat{V}(x,y,z)$ has been described phenomenologically by using a 3D harmonic-gaussian function representing a double quantum dot symmetric in the $x$ direction, and with a harmonic profile in the $y$ and $z$ directions. In particular, we have varied the parameters $A_2$ and $k$, which are related to the height and the width of the confinement potential internal barrier, respectively. 

We have initially established the energetic contribution along the $x$ direction for the first six quantum states of the system (n=0-5). We have analyzed the values of the parameter $A_2$ for which the energy values correspond to the degenerate and non-degenerate states. Regarding the $k$ parameter, we have highlighted the considerable increase in energy values when the values of this quantity tend to zero, increasing the confinement effects on the electron. As long as comparison was possible, we have obtained a good agreement with the values of energy as a function of $A_2$ and $k$ found in the literature.

We have obtained the entropy values $S^n_r$ as a function of $A_2$ and $k$ for the quantum states $n=0-5$. In the first situation, we conjecture that the entropy $S^n_r$ successfully maps the degeneration of states when we vary the coupling parameter $A_2$. We also conclude that information entropy, through $S^{0}_r$, is an indicator of the level of decoupling/coupling of the double quantum dot. Furthermore, taking into account that informational entropies are used as a measure of delocalization/localization of $\rho_x(x)$, we justify the fluctuations in the values of $S^n_r$ as a function of $A_2$ and present the study of the values of $S^n_r$ as a function of $k$.

In addition to the informational analysis, we have determined the values of $S^n_p$ and $S^n_t$ as functions of $A_2$ and $k$. In this treatment, analyzing trends and, through the derivative of $S^n_t$, we focus on general aspects of the behavior of the values obtained. Additionally, we conclude that all values obtained for $S^n_t$ respect the entropic uncertainty relationship. In future work we shall delve deeper into the physical explanations about the behavior of the values of $S^n_p$ and $S^n_t$ as a function of $A_2$ and $k$.

Finally, from another perspective of work, we shall also use Shannon's informational entropies to analyze an electron confined in a double quantum dot, however, this time subjected to external fields.\\

\textbf{Supporting Information}

This manuscript contains supplementary information. \href{https://ars.els-cdn.com/content/image/1-s2.0-S0921452624000334-mmc1.pdf}{{\color{blue}Click here to access.}} \\

\textbf{Acknowledgements}

The authors acknowledge Conselho Nacional de Desenvolvimento Científico e Tecnológico (CNPq) and Coordena\c{c}\~ao de Aperfei\c{c}oamento Pessoal de Nível Superior (CAPES) for the financial support. \\

\textbf{Conflict of Interest} 

The authors declare no conflict of interest.

\bibliographystyle{iopart-num}
\bibliography{RefQDots6}

\end{document}